\documentclass[12pt]{article}
\usepackage{amsmath,amssymb,epsfig}
\makeatletter \@addtoreset{equation}{section} \makeatother
\renewcommand{\theequation}{\thesection.\arabic{equation}}
\newcommand{\ba}{\begin{array}}
\newcommand{\ea}{\end{array}}
\newcommand{\beq}{\begin{equation}}
\newcommand{\eeq}{\end{equation}}
\newcommand{\bea}{\begin{eqnarray}}
\newcommand{\eea}{\end{eqnarray}}
\def\bce{\begin{center}}
\def\ece{\end{center}}

\def\nonu{\nonumber}

\def\pa{\partial}

\def\be{\beta}

\newcommand{\tr}{\mbox{Tr}}

\def\eps6{{\displaystyle \mathop{\epsilon}^{6}}{}}

\def\nab6{{\displaystyle \mathop{\nabla}^{6}}{}}
\def\0{{\sst{(0)}}}
\def\1{{\sst{(1)}}}
\def\2{{\sst{(2)}}}
\def\3{{\sst{(3)}}}
\def\4{{\sst{(4)}}}
\def\5{{\sst{(5)}}}
\def\6{{\sst{(6)}}}
\def\7{{\sst{(7)}}}
\def\8{{\sst{(8)}}}

\def\ba{\begin{array}}
\def\ea{\end{array}}
\def\beq{\begin{equation}}
\def\eeq{\end{equation}}
\def\be{\begin{equation}}
\def\ee{\end{equation}}

\def\tr{\mathop{\rm tr}}

\def\eps{\epsilon}

\def\ba{\begin{array}}
\def\ea{\end{array}}
\def\beq{\begin{equation}}
\def\eeq{\end{equation}}
\def\be{\begin{equation}}
\def\ee{\end{equation}}

\def\tr{\mathop{\rm tr}}

\def\eps{\epsilon}

\newcommand{\bean}{\begin{eqnarray*}}
\newcommand{\eean}{\end{eqnarray*}}

\begin{document}
\thispagestyle{empty} \addtocounter{page}{-1}
\begin{flushright}
%KIAS-P06003 \\
%CALT-68-nnnn \\
%{\tt hep-th/0701145}\\
\end{flushright}

\vspace*{1.3cm}

\centerline{ \Large \bf More on Meta-Stable Brane Configuration }
\vspace{.3cm} 
\centerline{ \Large \bf  by Quartic Superpotential for Fundamentals } 
\vspace*{1.5cm}
\centerline{{\bf Changhyun Ahn} 
%and {\bf Yutaka Ookouchi $^{2}$}
} 
\vspace*{1.0cm} 
\centerline{\it 
Department of Physics, Kyungpook National University, Taegu
702-701, Korea} 
%\centerline{\it $^{2}$ California Institute of 
%Technology, Pasadena, CA91125, USA }
\vspace*{0.8cm} 
\centerline{\tt ahn@knu.ac.kr} 
%\qquad
%yutaka@caltech.edu} 
\vskip2cm

\centerline{\bf Abstract}
\vspace*{0.5cm}

For the case where 
the gauge theory superpotential has 
a quartic term as well as the mass term for quarks,
the nonsupersymmetric meta-stable brane
configuration was found recently. 
By adding the orientifold 6-planes 
and the extra fundamental flavors to this brane
configuration,
we describe the meta-stable nonsupersymmetric 
vacua of the gauge theory with antisymmetric flavor
as well as fundamental flavors in type IIA string theory.  

\baselineskip=18pt
\newpage
\renewcommand{\theequation}
{\arabic{section}\mbox{.}\arabic{equation}}

%%%%%%%%%%%%%%%%%%%%%%%%%%%%%%%%%%%%%%%%%%%%%%%%%%%%%%%%%%%%%%%%%%%%%%%%%%
%%%%%%%%%%%%%%%%%%%%%%%%%%%%%%%%%%%%%%%%%%%%%%%%%%%%%%%%%%%%%%%%%%%%%%%%%%
\section{Introduction}
%%%%%%%%%%%%%%%%%%%%%%%%%%%%%%%%%%%%%%%%%%%%%%%%%%%%%%%%%%%%%%%%%%%%%%%%%%
%%%%%%%%%%%%%%%%%%%%%%%%%%%%%%%%%%%%%%%%%%%%%%%%%%%%%%%%%%%%%%%%%%%%%%%%%%

It is known that 
the dynamical supersymmetry breaking in meta-stable vacua \cite{ISS,IS} 
occurs 
in the standard ${\cal N}=1$ SQCD with massive fundamental 
flavors.
The extra mass term
for quarks in the superpotential implies that
some of the F-term equations cannot  be satisfied and then the
supersymmetry is broken.
The corresponding meta-stable brane realizations of type IIA string theory 
have been found in \cite{OO1,FGU,BGHSS}.  
Very recently Giveon and Kutasov \cite{GK0710-1,GK0710} have found 
the type IIA nonsupersymmetric meta-stable 
brane configuration where  an additional  quartic
term for quarks in the superpotential is present.
Geometrically, this extra deformation 
corresponds to the rotation of D6-branes 
along the (45)-(89) directions while keeping the other branes 
described in \cite{OO1,FGU,BGHSS} unchanged. 
Classically there exist only
supersymmetric ground states. 
By adding the orientifold 6-plane to this brane
configuration \cite{GK0710-1},
the meta-stable nonsupersymmetric 
vacua of the supersymmetric unitary gauge theory with symmetric flavor
plus fundamental flavors is found \cite{Ahn07-11}. 

Let us add an orientifold 6-plane and extra eight half 
D6-branes, 
located at the NS5'-brane, 
into the brane configuration of 
\cite{GK0710-1} together with an extra NS5-brane and the mirrors for
both D4-branes and rotated D6-branes. 
According to
the observation of \cite{LLL1,BHKL,EGKT}, this ``fork'' 
brane configuration
contains the NS5'-brane embedded in an O6-plane at $x^7=0$. This 
NS5'-brane divides the O6-plane into two separated regions corresponding
to positive $x^7$ and negative $x^7$. Then RR charge of the O6-plane
jumps from $-4$ to $+4$. Furthermore, eight semi-infinite D6-branes
are present in the positive $x^7$ region. This is necessary for the
vanishing of the six dimensional anomaly. 
Then 
the type IIA brane configuration consists of 
two NS5-branes, one NS5'-brane, 
D4-branes, rotated D6-branes,  an O6-plane and 
eight half D6-branes.
We'll see how the corresponding supersymmetric gauge theory, which is
a standard ${\cal N}=1$ SQCD with massive flavors together with the
extra matters, occurs in the context of dynamical supersymmetry
breaking in meta-stable vacua.

In this paper, we study ${\cal N}=1$ $SU(N_c)$ gauge theory with 
an antisymmetric flavor $A$, a conjugate symmetric flavor $\widetilde{S}$,
$N_f$ fundamental
flavors $Q$ and $\widetilde{Q}$ and eight fundamental flavors $\hat{Q}$
in the context of dynamical supersymmetric breaking vacua.
Now we deform this theory by  adding  both the mass
term and the quartic term for quarks $Q, \widetilde{Q}$ 
in the fundamental representation of the gauge group \cite{GK0710-1}. 
Then we turn to the
dual magnetic  gauge theory \cite{Ahn07-1}. 
The dual
magnetic theory giving rise to the meta-stable vacua 
is described by ${\cal N}=1$
$SU(2N_f-N_c+4)$ gauge theory with dual matter contents. 
The difference between the brane configuration of \cite{Ahn07-1}
and the brane configuration of this paper is that the D6-branes are 
rotated in the (45)-(89) directions. 
By analyzing the magnetic superpotential, along the line of 
\cite{GK0710-1,GK0710}, we present the behaviors of gauge theory
description and string theory description for the meta-stable vacua.

In section 2, the type IIA brane configuration corresponding
to the electric theory based on the ${\cal N}=1$ $SU(N_c)$ gauge theory
with above matter contents is given.
In section 3, we construct the Seiberg dual magnetic theory which is 
${\cal N}=1$ $SU(2N_f-N_c+4)$ gauge theory with corresponding dual
matters. The rotation of D6-branes is encoded in the
mass term for the meson field in the superpotential.
In section 4, the nonsupersymmetric meta-stable
minimum is found
and  
the corresponding intersecting brane configuration of type IIA string
theory is presented.
In section 5, we comment on the future works.

%%%%%%%%%%%%%%%%%%%%%%%%%%%%%%%%%%%%%%%%%%%%%%%%%%%%%%%%%%%%%%%%%%%%%%
%%%%%%%%%%%%%%%%%%%%%%%%%%%%%%%%%%%%%%%%%%%%%%%%%%%%%%%%%%%%%%%%%%%%%%
\section{The ${\cal N}=1$ supersymmetric electric brane configuration}
%%%%%%%%%%%%%%%%%%%%%%%%%%%%%%%%%%%%%%%%%%%%%%%%%%%%%%%%%%%%%%%%%%%%%%
%%%%%%%%%%%%%%%%%%%%%%%%%%%%%%%%%%%%%%%%%%%%%%%%%%%%%%%%%%%%%%%%%%%%%%

The type IIA supersymmetric electric
brane configuration \cite{LLL1,BHKL,EGKT,Ahn07-1} corresponding to 
${\cal N}=1$ $SU(N_c)$ gauge theory  with 
an  antisymmetric flavor $A$, a conjugate symmetric flavor
$\widetilde{S}$,
eight fundamental flavors $\hat{Q}$ and 
$N_f$ fundamental flavors $Q, \widetilde{Q}$ \cite{ILS}
can be described as follows: one middle NS5'-brane(012389), two
NS5-branes(012345) denoted by $NS5_L$-brane and $NS5_R$-brane 
respectively, 
$N_c$
D4-branes(01236)
between them, $2N_f$ D6-branes(0123789), an
orientifold 6 plane(0123789) of positive RR charge, 
an orientifold 6 plane(0123789) of negative RR charge and eight half
D6-branes. 
The transverse coordinates $(x^4, x^5, x^6)$ transform as $(-x^4, -x^5,
-x^6)$ under the orientifold 6-plane(O6-plane) action. 
Let us introduce two complex coordinates \cite{GK98}
\bea
v \equiv x^4 + i x^5, \qquad w \equiv x^8 + i x^9.
\nonu
\eea
Then the origin of
the coordinates $(x^6, v, w)$ is located at the intersection of
$x^6$ coordinate and O6-plane.
The left $NS5_L$-brane is located at the left hand side of O6-plane
while the right $NS5_R$-brane is located at the right hand side of 
O6-plane. The $N_c$ color D4-branes are suspended between 
$NS5_L$-brane and $NS5_R$-brane. Moreover the $N_f$ D6-branes 
are located between the $NS5_L$-brane and the middle NS5'-brane 
and its mirrors $N_f$ D6-branes  
are located between the middle NS5'-brane and the $NS5_R$-brane.
The antisymmetric and conjugate symmetric flavors $A$ and  
$\widetilde{S}$
are 4-4 strings stretching between D4-branes
located at the left hand side of O6-plane and those at the right hand
side of O6-plane,
$N_f$ fundamental flavors $Q$ and $\widetilde{Q}$  are strings
stretching between $N_f$ D6-branes and $N_c$ color D4-branes and 
eight fundamental flavors $\hat{Q}$   are strings
stretching between eight half 
D6-branes  which are on top of $O6^{-}$-plane 
and $N_c$ color D4-branes. 

Let us deform this theory which has vanishing superpotential
by adding both the mass term 
and  the quartic term for $N_f$ fundamental quarks. 
The former can be achieved by ``translating'' the D6-branes along $\pm v$
direction leading to their coordinates $v = \pm v_{D6}$ \cite{GK98} 
while the latter can be obtained by ``rotating'' the D6-branes
\cite{GK0710-1} 
by an angle 
$\theta$ in $(w,v)$-plane. We denote them by $D6_{\theta}$-branes
which are at angle $\theta$ with undeformed unrotated D6-branes(0123789).
Then their mirrors $N_f$ D6-branes are 
rotated by an angle $-\theta$ in $(w,v)$-plane according to 
O6-plane action and 
we denote them also by $D6_{-\theta}$-branes \footnote{Note that 
the convention for
$D6_{\theta}$-branes in \cite{Ahn07-1} was such that the angle between
unrotated D6-branes and $D6_{\theta}$-branes was not $\theta$ but 
$(\frac{\pi}{2}-\theta)$.}. 
Then, in the electric gauge theory, the deformed superpotential is
given by
\bea
W_{elec} & = & \frac{\alpha}{2} \tr (Q \widetilde{Q})^2 - m \tr Q
\widetilde{Q} -\frac{1}{2\mu} \left[ (A\widetilde{S})^2+ Q
  \widetilde{S} A \widetilde{Q} +(Q \widetilde{Q})^2 
\right] + \hat{Q} \widetilde{S} \hat{Q}, \nonu \\
 & & \mbox{with} \qquad \alpha = \frac{\tan \theta}{\Lambda},  
\qquad m = \frac{v_{D6}}{2\pi \ell_s^2}
\label{electricsuperpotential}
\eea 
where $\Lambda$ is related to the scales of the electric and magnetic 
theories and $\pm v_{D6}$ is the $v$ coordinate of 
$D6_{\mp \theta}$-branes.
Due to the last term, the flavor symmetry $SU(N_f+8)_L$ is broken to 
$SU(N_f)_L \times SO(8)_L$. Here the adjoint mass 
$\mu \equiv \tan (\frac{\pi}{2}-\omega)$
is related to a rotation angle $\omega$ 
of $NS5_{L,R}$-branes in $(w,v)$-plane. In the limit of $\mu
\rightarrow \infty$(or no rotations of NS5-branes 
$ \omega \rightarrow 0$), the terms of
$\frac{1}{\mu}$ in (\ref{electricsuperpotential}) vanish.

Let us summarize the ${\cal N}=1$ supersymmetric electric brane
configuration with nonvanishing superpotential 
(\ref{electricsuperpotential}) 
in type IIA string theory as follows and draw it in
Figure 1:

$\bullet$ Two NS5-branes in (012345) directions with $w=0$

$\bullet$ One NS5'-brane in (012389) directions with $v=0=x^6$

$\bullet$ $N_c$ color D4-branes in (01236) directions with $v=0=w$

$\bullet$ $N_f$ $D6_{\pm \theta}$-branes in (01237)
directions and
two other directions in $(v,w)$-plane 

$\bullet$ Eight half D6-branes in (0123789) directions with $x^6=0=v$

$\bullet$ $O6^{\pm}$-planes in (0123789) directions with $x^6=0=v$

By moving the $D6_{\pm \theta}$-branes from Figure 1 into the outside of 
$NS5_{L,R}$-branes, there exist $N_f$ flavor D4-branes
connecting $D6_{\pm \theta}$-branes and the $NS5_{L,R}$-branes, and  
the gauge singlet field $N$ appears. 
At energies much below the mass
of $N$, the two brane descriptions 
coincide with each other. 
One can think of this new brane configuration as integrating the field
$N$ in from Figure 1 and the superpotential of this electric theory
contains the interaction term between $N$ with electric quarks,
quadratic term  and linear term for $N$ \cite{GK0710-1}.
The classical supersymmetric vacua of this brane configuration are 
characterized by the parameter $k$ where $k=0,1, \cdots, N_c$ and
unbroken gauge symmetry in the $k$-th configuration is 
$SU(N_c-k)$. That is, the $k$ D4-branes among $N_f$ D4-branes 
(stretched between $NS5_R$-brane and  $D6_{-\theta}$-branes) are 
reconnecting with
those number of D4-branes stretched between the middle NS5'-brane and 
$NS5_R$-brane. Then those resulting 
$k$ D4-branes are moving to $\pm v$
direction and the remaining $(N_c-k)$ D4-branes are stretching 
between 
the middle NS5'-brane and $NS5_R$-brane and $(N_f-k)$ D4-branes are 
stretched between the $NS5_R$-brane and $D6_{-\theta}$-branes(and
their mirrors).

%%%%%%%%%%%%%%%%%%%%%%%%%%%%%%%%%%%%%%%%%%%%%%%%%%%%%%%%%%%%%%%%%%%%%%%
%%%%%%%%%%%%%%%%%%%%%%%%%%%%%%%%%%%%%%%%%%%%%%%%%%%%%%%%%%%%%%%%%%%%%%%
\begin{figure}[ht]
   \epsfxsize=3.5in 
\centerline{\epsffile{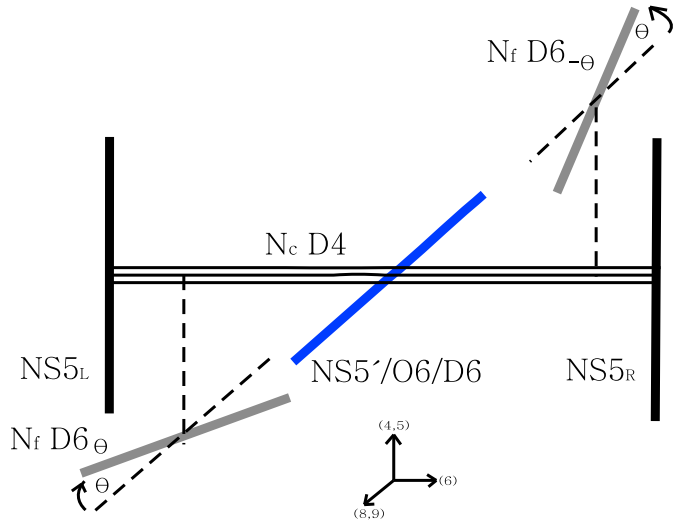}}
   \caption[FIG. \arabic{figure}.]{ 
The ${\cal N}=1$ 
supersymmetric electric brane configuration with deformed
superpotential (\ref{electricsuperpotential}) for the $SU(N_c)$ gauge
theory  with 
an antisymmetric flavor $A$, 
a conjugate symmetric flavor $\widetilde{S}$, eight 
fundamentals $\hat{Q}$,
and 
$N_f$ fundamental massive flavors $Q, \widetilde{Q}$. 
The origin of
the coordinates $(x^6, v, w)$ is located at the intersection of
$x^6$ and O6-plane. It is evident that the two deformations
are characterized by both translation and rotation for D6-branes. As
in \cite{Ahn07-1}, a combination of a middle
NS5'-brane, $O6^{+}$-plane, $O6^{-}$-plane and eight half D6-branes is
represented by $NS5'/O6/D6$.}
\end{figure}
%%%%%%%%%%%%%%%%%%%%%%%%%%%%%%%%%%%%%%%%%%%%%%%%%%%%%%%%%%%%%%%%%%%%%%%%%%%
%%%%%%%%%%%%%%%%%%%%%%%%%%%%%%%%%%%%%%%%%%%%%%%%%%%%%%%%%%%%%%%%%%%%%%%%%%%

%%%%%%%%%%%%%%%%%%%%%%%%%%%%%%%%%%%%%%%%%%%%%%%%%%%%%%%%%%%%%%%%%%%%% 
%%%%%%%%%%%%%%%%%%%%%%%%%%%%%%%%%%%%%%%%%%%%%%%%%%%%%%%%%%%%%%%%%%%%%%
\section{The ${\cal N}=1$ supersymmetric magnetic brane configuration}
%%%%%%%%%%%%%%%%%%%%%%%%%%%%%%%%%%%%%%%%%%%%%%%%%%%%%%%%%%%%%%%%%%%%%%
%%%%%%%%%%%%%%%%%%%%%%%%%%%%%%%%%%%%%%%%%%%%%%%%%%%%%%%%%%%%%%%%%%%%%%

The magnetic theory is obtained by interchanging
the $D6_{\pm \theta}$-branes and $NS5_{L,R}$-branes
while the linking number is preserved. 
After one moves the left $D6_{\theta}$-branes to the right all 
the way(and their
mirrors, right $D6_{-\theta}$-branes to the left) past the middle
NS5'-brane 
and
the right $NS5_R$-brane, the linking number counting \cite{Ahn07-1} 
implies that one should add $N_f$ D4-branes, corresponding to the meson
$M(\equiv Q \widetilde{Q})$, to the left 
side of all the right $N_f$ $D6_{\theta}$-branes(and their mirrors). 
Note that when a D6-brane crosses the middle NS5'-brane,
due to the parallelness of these, there was no creation of D4-branes.
That is, 
when the $D6_{\pm \theta}$-branes approach the middle NS5'-brane, one
should take $\theta =0$ limit(making 
D6-branes to be parallel to
the middle NS5'-brane) and then after they cross the
middle NS5'-brane, they return to the original positions 
given by $D6_{\pm \theta}$-branes as follows:     
\bea
D6_{\pm \theta}-\mbox{branes} \qquad \rightarrow \qquad 
 D6-\mbox{branes} \qquad \rightarrow \qquad
D6_{\pm \theta}-\mbox{branes}.
\label{condition}
\eea

Next, let us move the left $NS5_L$-brane to the right all the way past
O6-plane
(and its mirror, right $NS5_R$-brane to the left), and then 
the linking number
counting \cite{Ahn07-1} leads to the fact  
that the dual number of colors was $(2N_f-N_c+4)$.
There was a creation of D4-branes when the NS5-brane crosses an
O6-plane because they are not parallel to each other.
From this, the constant term 4 in the dual color above arises,
compared with the case of \cite{Ahn07}. 
Now we draw the magnetic brane configuration in Figure 2 where 
some of the flavor D4-branes are recombined with those of
$(2N_f-N_c+4)$ color D4-branes and  those combined resulting 
D4-branes are moved
into $\pm v$ direction.
One takes $k$ D4-branes from $N_f$ flavor D4-branes and reconnect them
to those from $(2N_f-N_c+4)$ color D4-branes in Figure 2 such that
the resulting branes are connecting from the $D6_{\theta}$-branes to
the $NS5_L$-brane directly. Their coordinates between
$D6_{\theta}$-branes and $k$ D4-branes will be $v=-v_{D6}$ in
order to minimize the energy.
This Figure 2 also can be obtained from the magnetic brane configuration of 
\cite{GK0710-1} by adding O6-planes and eight half D6-branes 
with the right presence of mirrors
under the O6-plane action.

%%%%%%%%%%%%%%%%%%%%%%%%%%%%%%%%%%%%%%%%%%%%%%%%%%%%%%%%%%%%%%%%%%%%%%%%%%%%
%%%%%%%%%%%%%%%%%%%%%%%%%%%%%%%%%%%%%%%%%%%%%%%%%%%%%%%%%%%%%%%%%%%%%%%%%%%%
\begin{figure}[ht]
   \epsfxsize=4.0in 
\centerline{\epsffile{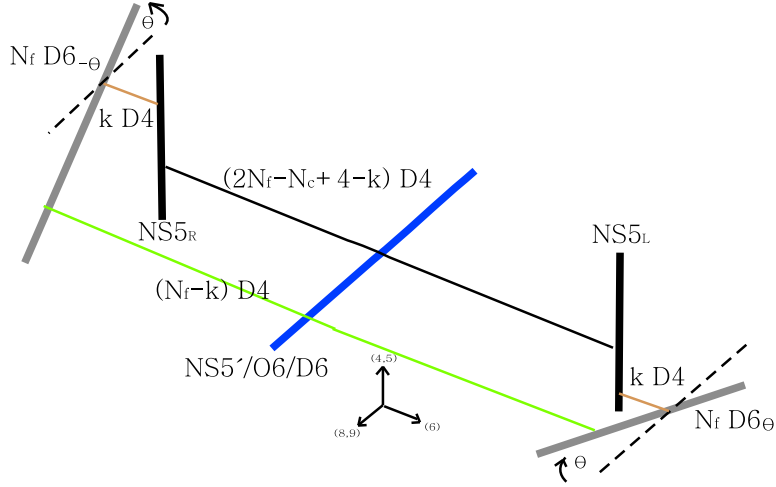}}
   \caption[FIG. \arabic{figure}.]{ 
The ${\cal N}=1$ supersymmetric magnetic 
brane configuration for the $SU(2N_f-N_c+4-k)$ gauge
theory  with 
an  antisymmetric flavor $a$, a conjugate symmetric flavor 
$\widetilde{s}$, 
$N_f$ fundamental flavors $q, \widetilde{q}$ and eight 
fundamentals $\hat{q}$.
The $N_f$ flavor D4-branes connecting between
$NS5_L$-brane and $D6_{\theta}$-branes are related to the dual gauge singlet
$M$ and are splitting into $(N_f-k)$ and
$k$ D4-branes. 
The location of intersection between $D6_{\theta}$-branes and $(N_f-k)$
D4-branes 
is given by $(v,w)=(0, v_{D6} \cot \theta)$ while the one between  
$D6_{\theta}$-branes and $k$
D4-branes 
is given by $(v,w)=(-v_{D6}, 0)$.   }
\end{figure}
%%%%%%%%%%%%%%%%%%%%%%%%%%%%%%%%%%%%%%%%%%%%%%%%%%%%%%%%%%%%%%%%%%%%%%%%%%%%% 
%%%%%%%%%%%%%%%%%%%%%%%%%%%%%%%%%%%%%%%%%%%%%%%%%%%%%%%%%%%%%%%%%%%%%%%%%%%%%

Then the low energy dynamics is described by the dual magnetic
theory with gauge group $SU(2N_f-N_c+4)$ and this theory is higgsed
down to  $SU(2N_f-N_c+4-k)$ in the $k$-th vacuum by nonzero vacuum
expectation value for dual quarks which is
determined later. The matters are 
$N_f$ flavors of fundamentals 
$q, \widetilde{q}$, an  antisymmetric flavor $a$, a conjugate symmetric
flavor $\widetilde{s}$, eight fundamentals $\hat{q}$, 
gauge singlet $M$ which is magnetic dual of
the electric meson field $Q \widetilde{Q}$ and other gauge singlet 
$\widetilde{M}$ that is $\hat{Q} \widetilde{Q}$.
Then the superpotential including the interaction between the meson
field $M$ and dual matters with $\mu \rightarrow \infty(\omega
\rightarrow 0) $ limit 
is described by
\bea
W_{mag} = \frac{1}{\Lambda} M q  \widetilde{s} a \widetilde{q} +  
\frac{\alpha}{2} \tr M^2- m \tr M + \hat{q} \widetilde{s} \hat{q} +
\widetilde{M} \hat{q} \widetilde{q}, \qquad 
M \equiv Q\widetilde{Q}, 
\qquad
\widetilde{M} \equiv \hat{Q} \widetilde{Q}
%\alpha = \frac{\tan \theta}{\Lambda}, 
%\qquad 
%m = \frac{v_{D6}}{2\pi \ell_s^2}
% +  \left[ (s \widetilde{s})^2+
%M_1 q \widetilde{q}+ P_0 q \widetilde{s} q +
%\widetilde{P}_0 \widetilde{q} s \widetilde{q} + \cdots \right] 
\label{mag}
\eea
where the second and third terms originate from 
the two deformations (\ref{electricsuperpotential}), and the fourth
term also comes 
 from electric theory. 
The $\theta$-dependent coefficient function, $\alpha$, in front of 
quadratic term of the meson field
also occurs in the geometric brane interpretation for the different
supersymmetric gauge theory \cite{Ahn06}. 
The $\alpha=0$ limit reduces to the theory 
given by \cite{Ahn07-1}. The parameters $\alpha$
and $m$ are the same as before in (\ref{electricsuperpotential})
 \footnote{For 
arbitrary rotation angles of $D6_{\pm \theta}$-branes and
$NS5_{\pm \omega}$-branes, there exist also other meson fields
containing $A$ or $\widetilde{S}$:$M_1
\equiv Q \widetilde{S} A \widetilde{Q}, P \equiv Q \widetilde{S} Q$ 
and $\widetilde{P}
\equiv \widetilde{Q} A \widetilde{Q}$. They couple to the dual quarks
and other flavors in the superpotential via $M_1 q \widetilde{q} +P q
\widetilde{s} q + \widetilde{P} \widetilde{q} a \widetilde{q}$. 
As emphasized in \cite{Ahn07-1},
the geometric constraint (\ref{condition}) at the intersection between
D6-branes and the middle NS5'-brane removes the
presence of these gauge singlets, $M_1, P$ and $\widetilde{P}$. That
is, when $D6_{\pm \theta}$-branes are crossing the middle NS5'-brane,
they do not produce any D4-branes because they are parallel at the
origin $x^6=0$. This implies there is no $M_1$ term in the magnetic 
superpotential. In general, $P$ and $\widetilde{P}$-terms arise when
$D6_{\theta}$-branes intersect with its mirrors $D6_{-\theta}$-branes
around the origin $x^6=0$. But they are also parallel to each other due to 
(\ref{condition}). This
leads to the fact 
that there are no $P$ or $\widetilde{P}$-terms in
the magnetic superpotential. Therefore, we are left with (\ref{mag}). }.

From the magnetic superpotential 
(\ref{mag}), the supersymmetric vacua are obtained and 
the F-term equations are given as follows:
\bea
\widetilde{s} a \widetilde{q} M & = & 0, \qquad a \widetilde{q} M q +
\hat{q} \hat{q} =0, 
\nonu \\
\widetilde{q} M q \widetilde{s} & = & 0, \qquad M q \widetilde{s}
a + \widetilde{M} \hat{q} =0,
\nonu \\
\widetilde{s} \hat{q} + \widetilde{q} \widetilde{M} & = & 0, \qquad
\hat{q} \widetilde{q} =0, \nonu \\
\frac{1}{\Lambda} q \widetilde{s} a \widetilde{q} & = & m -\alpha M. 
\label{Fterm}
\eea
By multiplying $M$ into the last equation with $\hat{q}=0=\widetilde{M}$,  
the matrix equation is satisfied
$
m M = \alpha M^2$.
Because the eigenvalues are either $0$ or $\frac{m}{\alpha}$, 
one can take
$N_f \times N_f$ matrix with $k$'s eigenvalues $0$ and $(N_f-k)$'s
eigenvalues $\frac{m}{\alpha}$:
\bea
M = \left(
\begin{array}{cc}
0 & 0  \\
0 & \frac{m}{\alpha} {\bf 1}_{N_f-k}
\end{array}
\right)
\label{M0}
\eea
where $k=1, 2, \cdots, N_f$ and ${\bf 1}_{N_f-k}$ is the $(N_f-k) \times
(N_f-k)$ identity matrix \cite{GK0710}.
The expectation value of $M$ is represented by 
the fundamental string between the flavor brane displaced by the 
$w$ direction and the color brane from Figure 2. 
The $k$ of the
$N_f$ flavor D4-branes are connected with $k$ of $(2N_f-N_c+4)$ color
D4-branes
and the resulting D4-branes stretch from the $D6_{\theta}$-branes to
the $NS5_L$-brane and the coordinate of an intersection point between the 
$k$ D4-branes and the $NS5_L$-brane is given by $(v, w)=(-v_{D6}, 0)$.
This corresponds to  exactly the $k$'s eigenvalues $0$ of $M$ above.
Now the remaining $(N_f-k)$ flavor D4-branes between 
the $D6_{\theta}$-branes and 
the NS5'-brane are related to the corresponding eigenvalues 
of $M$ given by $\frac{m}{\alpha} {\bf 1}_{N_f-k}$.
The coordinate of an intersection point between the 
$(N_f-k)$ D4-branes and the NS5'-brane is given 
by $(v, w)=(0, v_{D6} \cot \theta)$. Note that using the expressions
  for $\alpha$ and $m$ from (\ref{electricsuperpotential}), 
one obtains $\frac{m}{\alpha} =
\Lambda 
\frac{v_{D6} \cot \theta}{2\pi \ell_s^2}$ which must be 
$\Lambda \frac{w}{2\pi
  \ell_s^2}$. 
Then $w=v_{D6} \cot \theta$.

Substituting (\ref{M0}) into the last equation of (\ref{Fterm})
leads to 
\bea
q \widetilde{s} a \widetilde{q} = \left(
\begin{array}{cc}
m \Lambda {\bf 1}_k & 0  \\
0 & 0
\end{array}
\right).
\label{solqq}
\eea
Since the rank of the left hand side  is at most $2N_f-N_c+4$,
one must have more stringent bound $k \leq (2N_f-N_c+4)$.
In the $k$-th vacuum the gauge symmetry is broken to $SU(2N_f-N_c+4-k)$
and 
the supersymmetric vacuum drawn in Figure 2 with $k=0$ has 
$q \widetilde{s} = a \widetilde{q} =0$ and the gauge group 
$SU(2N_f-N_c+4)$ is unbroken. The expectation value of $M$ in this case 
is given by
$M = \frac{m}{\alpha} {\bf 1}_{N_f}= m \Lambda \cot 
\theta {\bf 1}_{N_f}$ explicitly.
By moving the D6-branes into the place between the middle NS5'-brane
and the $NS5_{L,R}$-branes, one obtains other brane configuration. 
There exist $(N_f-k)$ flavor D4-branes
connecting $D6_{\pm \theta}$-branes and the $NS5_{L,R}$-branes after
this movement. 

Another deformation arises when we rotate the $NS5_{L,R}$-branes by
an angle $\pm \theta'$ in the $(v,w)$-plane. This rotation provides an
adjoint field of $SU(2N_f-N_c+4)$ and couples to the magnetic dual
matters. Integrating this
adjoint field out leads to the fact that there exists 
a further contribution 
to the quartic 
superpotential  for $q$ and $\widetilde{q}$. In particular,
when the rotated $NS5_{\pm \theta'}$-branes are parallel to 
rotated $D6_{\pm \theta}$-branes, the coupling in front of 
$M^2$ in the magnetic
superpotential vanishes.

%%%%%%%%%%%%%%%%%%%%%%%%%%%%%%%%%%%%%%%%%%%%%%%%%%%%%%%%%%%%%%%%%%%%%%%%%%%%%
%%%%%%%%%%%%%%%%%%%%%%%%%%%%%%%%%%%%%%%%%%%%%%%%%%%%%%%%%%%%%%%%%%%%%%%%%%%%%
\section{Nonsupersymmetric meta-stable brane configuration }
%%%%%%%%%%%%%%%%%%%%%%%%%%%%%%%%%%%%%%%%%%%%%%%%%%%%%%%%%%%%%%%%%%%%%%%%%%%%%
%%%%%%%%%%%%%%%%%%%%%%%%%%%%%%%%%%%%%%%%%%%%%%%%%%%%%%%%%%%%%%%%%%%%%%%%%%%%%

The theory has many nonsupersymmetric meta-stable ground states
besides the supersymmetric ones we discussed in previous section.
For the IR free region \cite{Ahn07-1}, 
the magnetic theory is the effective low energy description of the
asymptotically free electric gauge theory.
When we rescale the meson field as
$
M = h \Lambda \Phi $,
then the Kahler potential for $\Phi$ is canonical and the magnetic
quarks are canonical near the origin of field space.
The higher order corrections of Kahler potential are negligible when 
the expectation values of the fields $q, \widetilde{q}, a,
\widetilde{s}$ 
and $\Phi$
are smaller than the scale of magnetic theory.
Then the magnetic superpotential can be written in terms of $\Phi$(or $M$)
\bea
W_{mag}  =  h \Phi  q  \widetilde{s} a \widetilde{q} 
 +  
\frac{h^2 \mu_{\phi}}{2} \tr \Phi^2- h \mu^2 \tr \Phi 
+ \hat{q} \widetilde{s} \hat{q} +
\widetilde{M} \hat{q} \widetilde{q}.
% \nonu \\
% & = &
% \frac{1}{\Lambda} M q  \widetilde{s} a \widetilde{q} +  
%\frac{\alpha}{2} \tr M^2- m \tr M + \hat{q} \widetilde{s} \hat{q} +
%\widetilde{M} \hat{q} \widetilde{q}.
\nonu
\eea
From this, one can read off the following quantities
\bea
\mu^2 = m \Lambda, 
%\left(= \frac{v_{D6}}{ g_s \ell_s^3} \right), 
\qquad 
\mu_{\phi} = \alpha \Lambda^2, 
%\left(= \frac{\tan \theta}{g_s    \ell_s} \right), 
\qquad M = h \Lambda \Phi.
\nonu
\eea

The classical supersymmetric vacua given by (\ref{M0}) and
(\ref{solqq})
can be described similarly 
%\bea
% h \Phi
% = \left(
%\begin{array}{cc}
%0 & 0  \\
%0 & \frac{\mu^2}{\mu_{\phi}} {\bf 1}_{N_f-k}
%\end{array}
%\right), \qquad
%q \widetilde{s} a \widetilde{q} = \left(
%\begin{array}{cc}
%\mu^2 {\bf 1}_k & 0  \\
%0 & 0
%\end{array}
%\right).
%\nonu
%\eea
and one decomposes 
the $(N_f-k) \times (N_f-k)$
block  at the lower right corner of $h\Phi$ and $q \widetilde{s} a
\widetilde{q}$ 
into blocks of 
size $n$ and $(N_f-k-n)$ as follows:
\bea
h\Phi = \left(
\begin{array}{ccc}
0 & 0 & 0  \\
0 & h \Phi_n & 0 \\
0 & 0 & \frac{\mu^2}{\mu_{\phi}} {\bf 1}_{N_f-k-n}
\end{array}
\right), \qquad
q \widetilde{s} a \widetilde{q} = \left(
\begin{array}{ccc}
\mu^2 {\bf 1}_k & 0 & 0  \\
0 & { \varphi} \widetilde{\beta} \gamma \widetilde{\varphi}  &  0 \\
0 & 0 & 0
\end{array}
\right).
\nonu
\eea
Here $\varphi$ and $\widetilde{\varphi}$ are $n \times (2N_f-N_c+4-k)$
dimensional matrices and correspond to $n$ flavors of fundamentals of
the gauge group $SU(2N_f-N_c+4-k)$ which is unbroken by the nonzero
expectation value of $q\widetilde{s}$ and $a \widetilde{q}$.
In Figure 3, 
they correspond to 
fundamental strings connecting the $n$ flavor D4-branes and
$(2N_f-N_c+4-k)$
color D4-branes.
This Figure 3 can be also obtained from the meta-stable brane configuration of 
\cite{GK0710-1} by adding O6-planes and eight half D6-branes  
with appropriate mirrors
under the O6-plane action.
The antisymmetric and conjugate symmetric flavors $\gamma$ and  
$\widetilde{\beta}$
are 4-4 strings stretching between $(2N_f-N_c+4-k)$ D4-branes
located at the left hand side of O6-plane and those at the right hand
side of O6-plane in Figure 3.
Both $\Phi_n$ and ${ \varphi} \widetilde{\beta} \beta
\widetilde{\varphi}$
are $n \times n$ matrices.
The supersymmetric ground state corresponds to
the vacuum expectation vaules given by 
$h\Phi_n= \frac{\mu^2}{\mu_{\phi}} {\bf 1}_{n}, 
\varphi \widetilde{\beta}=0=\gamma \widetilde{\varphi}$. 

%%%%%%%%%%%%%%%%%%%%%%%%%%%%%%%%%%%%%%%%%%%%%%%%%%%%%%%%%%%%%%%%%%%%%%%%%
%%%%%%%%%%%%%%%%%%%%%%%%%%%%%%%%%%%%%%%%%%%%%%%%%%%%%%%%%%%%%%%%%%%%%%%%%
\begin{figure}[ht]
   \epsfxsize=4.0in 
\centerline{\epsffile{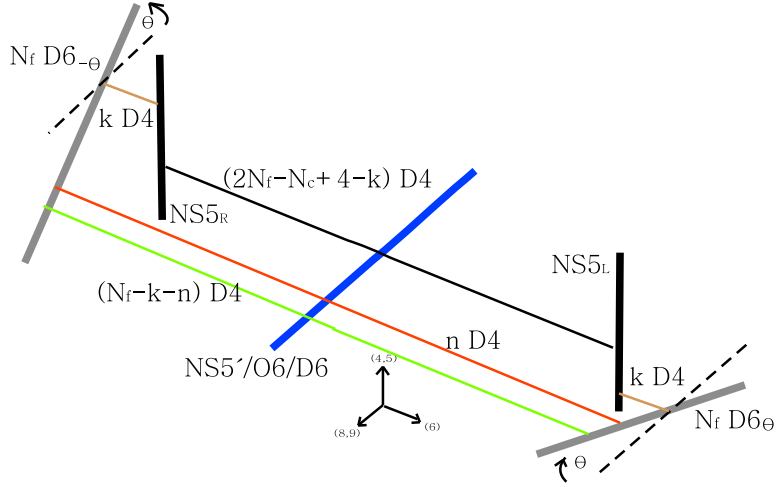}}
   \caption[FIG. \arabic{figure}.]{ 
The nonsupersymmetric minimal energy
brane configuration for the $SU(2N_f-N_c+4-k)$ gauge theory  with 
an antisymmetric flavor $a$, a conjugate symmetric flavor 
$\widetilde{s}$, 
$N_f$ fundamental flavors $q, \widetilde{q}$ and eight 
fundamentals $\hat{q}$.
This brane configuration is obtained by moving $n$ flavor D4-branes, 
from $(N_f-k)$ flavor D4-branes stretched
between the NS5'-brane and the $D6_{\theta}$-branes in Figure
2(and their mirrors).  }
\end{figure}
%%%%%%%%%%%%%%%%%%%%%%%%%%%%%%%%%%%%%%%%%%%%%%%%%%%%%%%%%%%%%%%%%%%%%%%%%%
%%%%%%%%%%%%%%%%%%%%%%%%%%%%%%%%%%%%%%%%%%%%%%%%%%%%%%%%%%%%%%%%%%%%%%%%%%

Now the full one loop potential for $\Phi_n, \hat{\varphi} 
\equiv \varphi \widetilde{\beta}, 
\hat{\widetilde{\varphi}} \equiv \gamma \widetilde{\varphi} $
\cite{Ahn07-1} 
takes the form
\bea
\frac{V}{|h|^2}  =  
|\Phi_n  \hat{\varphi}|^2   
+  |\Phi_n  \hat{\widetilde{\varphi}}  |^2
  +  
| \hat{\varphi}  \hat{\widetilde{\varphi}}-\mu^2 {\bf 1}_{n} + 
h \mu_{\phi} \Phi_n|^2 + b |h \mu|^2 \tr \Phi_n^{\dagger} \Phi_n, 
\nonu
\eea
where $b$ is given by
$b = \frac{(\ln 4-1)}{8\pi^2} (2N_f-N_c+4)$.
Differentiating this potential with respect to 
$\Phi_n$ and putting $\hat{\varphi}=0=\hat{
\widetilde{\varphi}}$, one obtains
\bea
h \Phi_n = \frac{\mu^2 \mu_{\phi}^{\ast}}{|\mu_{\phi}|^2 +b |\mu|^2}
{\bf 1}_n
\simeq \frac{\mu^2 \mu_{\phi}^{\ast}}{b |\mu|^2}
{\bf 1}_n \qquad \mbox{or} \qquad
M_n \simeq \frac{\alpha \Lambda^3}{(2N_f-N_c+4)} {\bf 1}_{n}
\label{vac}
\eea
for real $\mu$ and 
we assume here that 
$\mu_{\phi} << \mu << \Lambda_m$. The vacuum energy $V$ is given by
$V \simeq n |h \mu^2|^2$.
Expanding around this solution, one obtains
the eigenvalues for mass matrix for $\hat{\varphi}$ and 
$\hat{\widetilde{\varphi}}$ 
will be
\bea
m_{\pm}^2 = \frac{|\mu|^4}{(|\mu_{\phi}|^2 + b |\mu|^2)^2}\left[ 
|\mu_{\phi}|^2 \pm b |h|^2 \left( |\mu_{\phi}|^2 + b|\mu|^2 \right)
\right]
\simeq \frac{1}{b^2} \left( |\mu_{\phi}|^2  \pm |b h \mu|^2 \right).
\nonu
\eea
Then  for 
$
 | \frac{\mu_{\phi}}{\mu}  |^2 
> \frac{|b h|^2}{1-b|h|^2} \simeq |bh|^2
$ in order to avoid tachyons
the vacuum (\ref{vac}) is locally stable.

One can move $n$ D4-branes, from $(N_f-k)$ D4-branes stretched
between the NS5'-brane and the $D6_{\theta}$-branes at $w=v_{D6}
\cot \theta $, to the local minimum of the potential and the end
points of these $n$ D4-branes are at a nonzero $w$ as in Figure 3.
The remaining $(N_f-k-n)$ flavor D4-branes between 
the $D6_{\theta}$-branes and 
the NS5'-brane are related to the corresponding eigenvaules 
of $h\Phi$,   i.e., $\frac{\mu^2}{\mu_{\phi}} {\bf 1}_{N_f-k-n}$.
The coordinate of an intersection point between the 
$(N_f-k-n)$ D4-branes and the NS5'-brane is given 
by $(v, w)=(0, v_{D6} \cot \theta)$.
Finally, 
the remnant $n$ ``curved'' flavor D4-branes between 
the $D6_{\theta}$-branes and 
the NS5'-brane are related to the corresponding eigenvaules 
of $h\Phi_n$ by (\ref{vac}).
Note that since $  \frac{\mu^2 
\mu_{\phi}^{\ast}}{b |\mu|^2} << \frac{\mu^2}{\mu_{\phi}}$,
the $n$ D4-branes are nearer to the $w=0$ located at the $NS5_L$-brane.

As explained in \cite{GK0710-1}, this local stable vacuum decays to
the supersymmetric ground states. The end points of $n$ ``curved'' flavor
D4-branes
on the NS5'-brane approach those  of the $(2N_f-N_c+4-k)$ color 
D4-branes and two types of branes reconnect each other. 
For $n \leq (2N_f-N_c+4-k)$, the final brane configuration is nothing
but the supersymmetric vacuum of Figure 2 with the replacement 
$k \rightarrow (k+n)$.
When $n > (2N_f-N_c+4-k)$, then the remnant $[n-(2N_f-N_c+4-k)]$ 
flavor D4-branes remain.   
On the other hand, the $n$ D4-branes can move to larger $w$ and return
to the Figure 2. Also some of the D4-branes approach the intersection
point between $D6_{\theta}$-branes and the NS5'-brane while the
remaining D4-branes move to the one between $D6_{\theta}$-branes and
the $NS5_L$-brane.  

When $D6_{\pm \theta}$-branes are moved to the place between 
the NS5'-brane and the $NS5_{L,R}$-branes,
one gets the Figure 4 where the previous $(N_f-k)$ D6-branes in Figure
3 that were
not connected to the $NS5_{L,R}$-branes, through the flavor D4-branes,
are now connecting to the $NS5_{L,R}$-branes by the same number of 
D4-branes while  the previous $k$ D6-branes in Figure
3 that were
connected to the $NS5_{L,R}$-branes, through the flavor D4-branes,
are now not connecting to the $NS5_{L,R}$-branes.  
The former corresponds to a creation of D4-branes and the latter
corresponds to an annihilation of D4-branes due to the Hanany-Witten 
transition \cite{HW,GK98}.

%%%%%%%%%%%%%%%%%%%%%%%%%%%%%%%%%%%%%%%%%%%%%%%%%%%%%%%%%%%%%%%%%%%%%%%%%%%%
%%%%%%%%%%%%%%%%%%%%%%%%%%%%%%%%%%%%%%%%%%%%%%%%%%%%%%%%%%%%%%%%%%%%%%%%%%%%
\begin{figure}[ht]
   \epsfxsize=4.0in 
\centerline{\epsffile{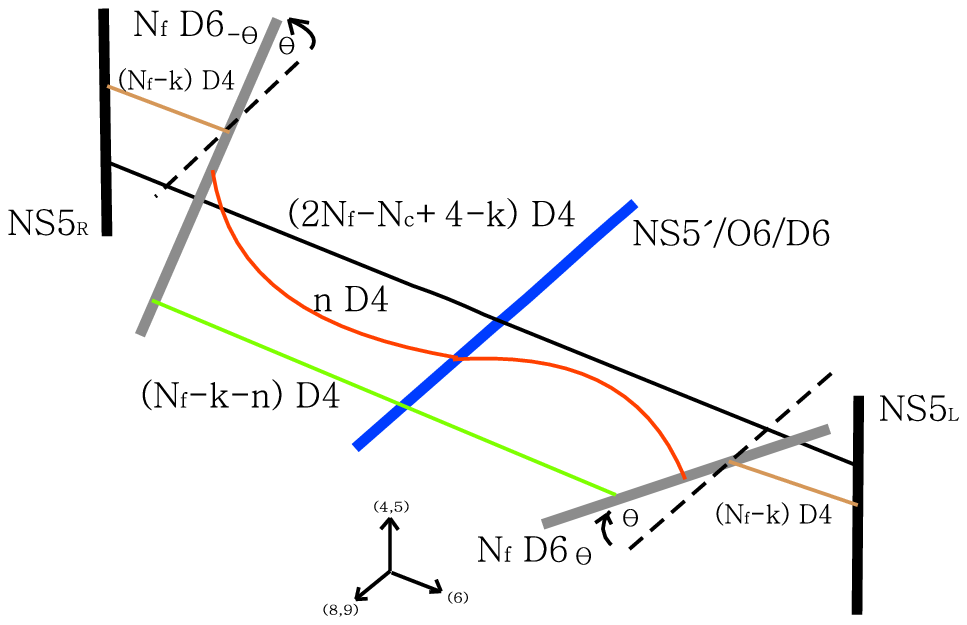}}
   \caption[FIG. \arabic{figure}.]{ 
The nonsupersymmetric minimal energy
brane configuration for the $SU(2N_f-N_c+4-k)$ gauge theory  with 
an antisymmetric flavor $a$, a conjugate symmetric 
flavor $\widetilde{s}$, 
$N_f$ fundamental flavors $q, \widetilde{q}$ and eight 
fundamentals $\hat{q}$.
This brane configuration can be obtained by 
moving $N_f$ $D6_{\pm \theta}$-branes
into the place between the $NS5_{L,R}$-branes and the middle 
NS5'-brane in Figure
3(and their mirrors). Note that there exists a creation of $(N_f-k)$
D4-branes connecting the $D6_{\theta}$-branes 
and the $NS5_{L}$-brane(and their mirrors). }
\end{figure}
%%%%%%%%%%%%%%%%%%%%%%%%%%%%%%%%%%%%%%%%%%%%%%%%%%%%%%%%%%%%%%%%%%%%%%%%%%%%%
%%%%%%%%%%%%%%%%%%%%%%%%%%%%%%%%%%%%%%%%%%%%%%%%%%%%%%%%%%%%%%%%%%%%%%%%%%%%%

In the remaining part, 
we focus on the gravitational potential of the $NS5_L$-brane.
Let us remind that the branes are placed as follows:
\bea
D6_{\theta}-\mbox{branes}(01237vw) & : & \qquad v  =  -v_{D6} +  w \tan
\theta, \nonu \\
NS5'-\mbox{brane}(012389) & : & \qquad  v  =  0, \nonu \\
NS5_{L}-\mbox{brane}(012345) & : & \qquad w  =  0
\nonu
\eea
where we assume that $D6_{\theta}$-branes are located at the nonzero 
\bea
y
\equiv x^6
\nonu 
\eea 
and 
the $NS5'$-brane is located at $y=0$(the origin).
The dependence of the distance between the $D6_{\theta}$-branes and 
$NS5'$ brane along the $NS5_L$-brane on $w$ can be represented by
\bea
\Delta x =|-v_{D6} +  w \tan \theta |.
\label{wdependence}
\eea
Then the partial differentiation of $\Delta x$ with
respect to
$w$ leads to an extra contribution for the computation of 
$\pa_w \theta_{i,w}$ where we define $\theta_{i,w}$ as 
\bea
\theta_{1,w} \equiv
\cos^{-1} \left(\frac{y_m}{|w|}
  \right), \qquad 
\theta_{2,w} \equiv
\cos^{-1} \left(\frac{y_m}{\sqrt{y^2+|w|^2}}
  \right)
\label{thetaw}
\eea
with $y_m$ that is smallest value of $y$ along the D4-brane. 
Then the energy density of the D4-brane was computed in 
\cite{GK0703,Ahn07-5}
and is given by
\bea
E(w) = \frac{\tau_4}{2 \ell_s} \sqrt{1+
\frac{\ell_s^2}{y_m^2}} \left[ 
|w|^2 \sin 2\theta_{1,w} + (y^2 +|w|^2) \sin 2\theta_{2,w} \right]
\label{energy}
\eea
corresponding to (3.7) of \cite{GK0710-1} 
where $\tau_4$ is a tension of D4-brane in flat spacetime.

It is straightforward to compute 
the differentiation of $  \left( \frac{\ell_s E(w)}{\tau_4} \right)^2$
with respect to $w$ and it leads to
\bea
&& \pa_w   ( \frac{\ell_s E(w)}{\tau_4} )^2  = 
 \ell_s^2 \bar{w} \left( \sin^2 \theta_{1,w} +  
\sin^2 \theta_{2,w} + [ 
\frac{\sqrt{y^2+|w|^2}}{|w|}  +  \frac{|w|}{\sqrt{y^2+|w|^2}} ]
\sin \theta_{1,w} \sin \theta_{2,w} \right) \nonu \\
&& + \left[ |w|^2 \sin 2 \theta_{1,w} +\left(y^2+|w|^2\right) 
\sin 2 \theta_{2,w}
\right]
 \times  \left[ \left(\ell_s^2 + |w|^2 \cos 2 \theta_{1,w} \right)
  \pa_w \theta_{1,w} \right. \nonu \\
&& \left. + \left(\ell_s^2 + \left(y^2+|w|^2 \right) \cos 2 
\theta_{2,w} \right)
  \pa_w \theta_{2,w} + \frac{\bar{w}}{2} \left( \sin 2 \theta_{1,w} +
\sin 2 \theta_{2,w} \right)  \right].
\label{equ}
\eea
In order to simplify this, one uses the partial differentiation of
$\Delta x$ (\ref{wdependence}) with respect to $w$ which is 
equal to $\frac{1}{2} 
\tan \theta  \frac{\bar{w} \bar{\tan \theta}  -\bar{v_{D6}}}{|w \tan
  \theta  -v_{D6}|}$. On the other hand, $\Delta x$ was known in
\cite{GK0703}
and it is 
\bea
\Delta x = \frac{1}{2\ell_s} \left[ 
|w|^2 \sin 2\theta_{1,w} + (y^2 +|w|^2) \sin 2\theta_{2,w} \right]  
+ \ell_s \left(
  \theta_{1,w}+\theta_{2,w}
\right).
\label{dx}
\eea
After differentiating this (\ref{dx}) with respect to $w$ 
and equating it to the
previous expression obtained from (\ref{wdependence}), one arrives at 
\bea
&& 
\left(\ell_s^2 + |w|^2 \cos 2 \theta_{1,w} \right)
  \pa_w \theta_{1,w} + 
\left[\ell_s^2 + \left(y^2+|w|^2 \right) \cos 2 \theta_{2,w} \right]
  \pa_w \theta_{2,w} \nonu \\
&& + \frac{\bar{w}}{2} \left( \sin 2 \theta_{1,w} +
\sin 2 \theta_{2,w} \right)
  = \ell_s \frac{1}{2} 
\tan \theta  \frac{\bar{w} \bar{\tan \theta}  -\bar{v_{D6}}}{|w \tan
  \theta  -v_{D6}|} 
\label{relat}
\eea
corresponding to (3.21) of \cite{GK0710-1}.
Now by putting this relation (\ref{relat}) into 
the second and third lines of (\ref{equ}), one gets 
\bea
\pa_w 
 ( \frac{\ell_s E(w)}{\tau_4} )^2  & = &
 \ell_s^2 \bar{w} \left( \sin^2 \theta_{1,w} +  
\sin^2 \theta_{2,w} + [ 
\frac{\sqrt{y^2+|w|^2}}{|w|}  +  \frac{|w|}{\sqrt{y^2+|w|^2}} ]
\sin \theta_{1,w} \sin \theta_{2,w} \right)
\nonu \\
& + & 
 \left[ |w|^2 \sin 2 \theta_{1,w} +\left(y^2+|w|^2\right) \sin 2 \theta_{2,w}
\right] \frac{1}{2}
\ell_s \tan \theta   \frac{\bar{w} \bar{\tan \theta}
  -\bar{v_{D6}}}{|w \tan
  \theta  -v_{D6}|}.
\label{rel}
\eea
It is easy to see that at $w = v_{D6} \cot \theta$ which is an
intersection point between $D6_{\theta}$-branes and $NS5'$-brane, 
$\Delta x$ vanishes through (\ref{wdependence}) and this also implies 
that $\theta_{i,w}$ vanishes from (\ref{dx}). Furthermore, 
the energy 
$E(w)$ is zero from (\ref{energy}). This corresponds to the global
minimal energy. 
For the parallel $D6$-branes and NS5'-brane(i.e.,
$\tan \theta =0$), then the only stationary point is $w=0$
\cite{Ahn07-1}. If $w \neq 0$, then 
$\sin \theta_{1,w}=0=\sin \theta_{2,w}$ from 
the first term of (\ref{rel}) but these are
not physical solutions. 
 
For real and positive parameters $v_{D6}, w$ and $\tan \theta$, we are
looking for the solution with $v_{D6} > w \tan \theta $
and setting the right hand side of (\ref{rel}) to zero,
finally one gets with (\ref{thetaw})
\bea
\frac{
 \sin^2 \theta_{1,w} +  
\sin^2 \theta_{2,w} + \left( 
\frac{\sqrt{y^2+w^2}}{w}  +  \frac{w}{\sqrt{y^2+w^2}} \right)
\sin \theta_{1,w} \sin \theta_{2,w} }{
  w^2 \sin 2 \theta_{1,w} +\left(y^2+w^2\right) \sin 2 \theta_{2,w}
 } 
 = \frac{\tan \theta }{2 \ell_s w}.
\label{relation}
\eea
Therefore, the brane configuration of Figure 4 has a local minimum
where the end of D4-brane are located at $w$ given by (\ref{relation}).
When the $y$ goes to zero$(\theta_{1,w} = \theta_{2,w} \equiv 
\theta_{w})$, one can approximate 
(\ref{relation}) and one gets
\bea
\tan \theta_{w}  \simeq \frac{ w \tan \theta}{2\ell_s}.
%\label{wvalue}
\nonu
\eea
The gauge theory result is valid only when $\theta$ and
$\frac{v_{D6}}{\ell_s}$
are much smaller than $g_s$ while the classical brane construction with
(\ref{relation})
is valid for any angle and the length parameters are of order $\ell_s$
or larger. 

%%%%%%%%%%%%%%%%%%%%%%%%%%%%%%%%%%%%%%%%%%%%%%%%%%%%%%%%%%%%%%%%%%%%%%%%%%%%%
%%%%%%%%%%%%%%%%%%%%%%%%%%%%%%%%%%%%%%%%%%%%%%%%%%%%%%%%%%%%%%%%%%%%%%%%%%%%%
\section{Conclusions and outlook}
%%%%%%%%%%%%%%%%%%%%%%%%%%%%%%%%%%%%%%%%%%%%%%%%%%%%%%%%%%%%%%%%%%%%%%%%%%%%
%%%%%%%%%%%%%%%%%%%%%%%%%%%%%%%%%%%%%%%%%%%%%%%%%%%%%%%%%%%%%%%%%%%%%%%%%%%%

In this paper, by adding the orientifold 6-planes 
and the extra fundamental flavors to the brane
configuration \cite{GK0710-1},
we have described the meta-stable nonsupersymmetric 
vacua of the gauge theory with antisymmetric flavor
as well as fundamental flavors, through the Figures 3 and 4, 
from type IIA string theory.

It would be interesting to deform the theories given in 
\cite{Ahn07-4,Ahn07-5,Ahn07-7,Ahn07-9,Ahn07-10} where one of the gauge
group factor has the same matter contents as the one of the present paper 
and see how the meta-stable ground states appear.
Along the lines of 
\cite{GK0703,Ahn07-5,Ahn07-6,Ahn07-7,Ahn07-10}, when the
$D6_{\theta}$-branes
are replaced by a single $NS5_{\theta}$-brane, it would be interesting to see how
the present deformation arises in these theories.  
It is also possible to deform the symplectic or orthogonal gauge
group theory with massive flavors 
\cite{Ahn06-1} by adding an orientifold 4-plane. 
Similar application to the product gauge group
case \cite{Ahn07-2,Ahn07-3,Ahn07-8} is also possible to study.
It is an open problem to see how the type IIB description
\cite{TW0711} is related to the present work. To construct a direct
gauge mediation \cite{HM,XY} for the present work is also possible 
open problem.

\vspace{.7cm}

%%%%%%%%%%%%%%%%%%%%%%%%%%%%%%%%%%
\centerline{\bf Acknowledgments}
%%%%%%%%%%%%%%%%%%%%%%%%%%%%%%%%%%

We would like to thank D. Kutasov for discussions.
This work was supported by grant No.
R01-2006-000-10965-0 from the Basic Research Program of the Korea
Science \& Engineering Foundation.


\begin{thebibliography}{99}

\bibitem{ISS}
  K.~Intriligator, N.~Seiberg and D.~Shih,
  ``Dynamical SUSY breaking in meta-stable vacua,''
  JHEP {\bf 0604}, 021 (2006)
  [arXiv:hep-th/0602239].
  %%CITATION = JHEPA,0604,021;%%

%\cite{Intriligator:2007cp}
\bibitem{IS}
  K.~Intriligator and N.~Seiberg,
  ``Lectures on Supersymmetry Breaking,''
  Class.\ Quant.\ Grav.\  {\bf 24}, S741 (2007)
  [arXiv:hep-ph/0702069].
  %%CITATION = CQGRD,24,S741;%%

%\cite{Ooguri:2006bg}
\bibitem{OO1}
  H.~Ooguri and Y.~Ookouchi,
  ``Meta-stable supersymmetry breaking vacua on intersecting branes,''
  Phys.\ Lett.\ B {\bf 641}, 323 (2006)
  [arXiv:hep-th/0607183].
  %%CITATION = HEP-TH 0607183;%%

%\cite{Franco:2006ht}
\bibitem{FGU}
  S.~Franco, I.~Garcia-Etxebarria and A.~M.~Uranga,
  ``Non-supersymmetric meta-stable vacua from brane configurations,''
  JHEP {\bf 0701}, 085 (2007)
  [arXiv:hep-th/0607218].
  %%CITATION = HEP-TH 0607218;%%

%\cite{Bena:2006rg}
\bibitem{BGHSS}
  I.~Bena, E.~Gorbatov, S.~Hellerman, N.~Seiberg and D.~Shih,
  ``A note on (meta)stable brane configurations in MQCD,''
  JHEP {\bf 0611}, 088 (2006)
  [arXiv:hep-th/0608157].
  %%CITATION = HEP-TH 0608157;%%

%\cite{Giveon:2007ew}
\bibitem{GK0710-1}
  A.~Giveon and D.~Kutasov,
  ``Stable and Metastable Vacua in Brane Constructions of SQCD,''
  arXiv:0710.1833 [hep-th].
  %%CITATION = ARXIV:0710.1833;%%

%\cite{Giveon:2007ef}
\bibitem{GK0710}
  A.~Giveon and D.~Kutasov,
  ``Stable and Metastable Vacua in SQCD,''
  arXiv:0710.0894 [hep-th].
  %%CITATION = ARXIV:0710.0894;%%

%\cite{Ahn:2007si}
\bibitem{Ahn07-11}
  C.~Ahn,
  ``Other Meta-Stable Brane Configuration by Adding an Orientifold 6-Plane to
  Giveon-Kutasov,''
  arXiv:0712.0032 [hep-th].
  %%CITATION = ARXIV:0712.0032;%%

%\cite{Landsteiner:1998gh}
\bibitem{LLL1}
  K.~Landsteiner, E.~Lopez and D.~A.~Lowe,
  ``Duality of chiral N = 1 supersymmetric gauge theories via branes,''
  JHEP {\bf 9802}, 007 (1998)
  [arXiv:hep-th/9801002].
  %%CITATION = HEP-TH 9801002;%%

%\cite{Brunner:1998jr}
\bibitem{BHKL}
  I.~Brunner, A.~Hanany, A.~Karch and D.~Lust,
  ``Brane dynamics and chiral non-chiral transitions,''
  Nucl.\ Phys.\ B {\bf 528}, 197 (1998)
  [arXiv:hep-th/9801017].
  %%CITATION = HEP-TH 9801017;%%

%\cite{Elitzur:1998ju}
\bibitem{EGKT}
  S.~Elitzur, A.~Giveon, D.~Kutasov and D.~Tsabar,
  ``Branes, orientifolds and chiral gauge theories,''
  Nucl.\ Phys.\ B {\bf 524}, 251 (1998)
  [arXiv:hep-th/9801020].
  %%CITATION = HEP-TH 9801020;%%

%\cite{Ahn:2007uu}
\bibitem{Ahn07-1}
  C.~Ahn,
  ``More on Meta-Stable Brane Configuration,''
  Class.\ Quant.\ Grav.\  {\bf 24}, 3603 (2007)
  [arXiv:hep-th/0702038].
  %%CITATION = CQGRD,24,3603;%%

%\cite{Intriligator:1995ax}
\bibitem{ILS}
  K.~A.~Intriligator, R.~G.~Leigh and M.~J.~Strassler,
  ``New examples of duality in chiral and nonchiral supersymmetric gauge
  theories,''
  Nucl.\ Phys.\ B {\bf 456}, 567 (1995)
  [arXiv:hep-th/9506148].
  %%CITATION = HEP-TH 9506148;%%

%\cite{Giveon:1998sr}
\bibitem{GK98}
  A.~Giveon and D.~Kutasov,
  ``Brane dynamics and gauge theory,''
  Rev.\ Mod.\ Phys.\  {\bf 71}, 983 (1999)
  [arXiv:hep-th/9802067].
  %%CITATION = HEP-TH 9802067;%%

%\cite{Ahn:2007eh}
\bibitem{Ahn07}
  C.~Ahn,
  ``Meta-stable brane configuration with orientifold 6 plane,''
  JHEP {\bf 0705}, 053 (2007)
  [arXiv:hep-th/0701145].
  %%CITATION = JHEPA,0705,053;%%

%\cite{Ahn:2006gn}
\bibitem{Ahn06}
  C.~Ahn,
  ``Brane configurations for nonsupersymmetric meta-stable vacua in SQCD with
  adjoint matter,''
  Class.\ Quant.\ Grav.\  {\bf 24}, 1359 (2007)
  [arXiv:hep-th/0608160].
  %%CITATION = CQGRD,24,1359;%%

%\cite{Hanany:1996ie}
\bibitem{HW}
  A.~Hanany and E.~Witten,
  ``Type IIB superstrings, BPS monopoles, and three-dimensional gauge
  dynamics,''
  Nucl.\ Phys.\ B {\bf 492}, 152 (1997)
  [arXiv:hep-th/9611230].
  %%CITATION = HEP-TH 9611230;%%

%\cite{Giveon:2007fk}
\bibitem{GK0703}
  A.~Giveon and D.~Kutasov,
  ``Gauge symmetry and supersymmetry breaking from intersecting branes,''
  Nucl.\ Phys.\  B {\bf 778}, 129 (2007)
  [arXiv:hep-th/0703135].
  %%CITATION = NUPHA,B778,129;%%

%\cite{Ahn:2007um}
\bibitem{Ahn07-5}
  C.~Ahn,
  ``Meta-Stable Brane Configurations by Adding an Orientifold-Plane to
  Giveon-Kutasov,''
  JHEP {\bf 0708}, 021 (2007)
  [arXiv:0706.0042 [hep-th]].
  %%CITATION = JHEPA,0708,021;%%

%\cite{Ahn:2007ym}
\bibitem{Ahn07-4}
  C.~Ahn,
  ``Meta-stable brane configurations with five NS5-branes,''
  arXiv:0705.0056 [hep-th].
  %%CITATION = ARXIV:0705.0056;%%

%\cite{Ahn:2007cg}
\bibitem{Ahn07-7}
  C.~Ahn,
  ``Meta-Stable Brane Configurations with Seven NS5-Branes,''
  arXiv:0708.0439 [hep-th].
  %%CITATION = ARXIV:0708.0439;%%

%\cite{Ahn:2007gb}
\bibitem{Ahn07-9}
  C.~Ahn,
  ``Meta-Stable Brane Configurations of Multiple Product Gauge Groups with
  Orientifold 6 Plane,''
  arXiv:0710.0180 [hep-th].
  %%CITATION = ARXIV:0710.0180;%%

%\cite{Ahn:2007vz}
\bibitem{Ahn07-10}
  C.~Ahn,
  ``Meta-Stable Brane Configurations with Multiple NS5-Branes,''
  arXiv:0711.0082 [hep-th].
  %%CITATION = ARXIV:0711.0082;%%

%\cite{Ahn:2007yq}
\bibitem{Ahn07-6}
  C.~Ahn,
  ``More Meta-Stable Brane Configurations without D6-Branes,''
  Nucl.\ Phys.\  B {\bf 790}, 281 (2008)
  [arXiv:0707.0092 [hep-th]].
  %%CITATION = NUPHA,B790,281;%%

%\cite{Ahn:2006tg}
\bibitem{Ahn06-1}
  C.~Ahn,
  ``M-theory lift of meta-stable brane configuration in symplectic and
  orthogonal gauge groups,''
  Phys.\ Lett.\  B {\bf 647}, 493 (2007)
  [arXiv:hep-th/0610025].
  %%CITATION = PHLTA,B647,493;%%

%\cite{Ahn:2007er}
\bibitem{Ahn07-2}
  C.~Ahn,
  ``Meta-Stable Brane Configuration and Gauged Flavor Symmetry,''
  Mod.\ Phys.\ Lett.\  A {\bf 22}, 2329 (2007)
  [arXiv:hep-th/0703015].
  %%CITATION = MPLAE,A22,2329;%%

%\cite{Ahn:2007ve}
\bibitem{Ahn07-3}
  C.~Ahn,
  ``Meta-Stable Brane Configuration of Product Gauge Groups,''
  arXiv:0704.0121 [hep-th].
  %%CITATION = ARXIV:0704.0121;%%

%\cite{Ahn:2007gz}
\bibitem{Ahn07-8}
  C.~Ahn,
  ``Meta-Stable Brane Configurations of Triple Product Gauge Groups,''
  arXiv:0708.4255 [hep-th].
  %%CITATION = ARXIV:0708.4255;%%

%\cite{Tatar:2007wp}
\bibitem{TW0711}
  R.~Tatar and B.~Wetenhall,
  ``SQCD Vacua and Geometrical Engineering,''
  arXiv:0711.2534 [hep-th].
  %%CITATION = ARXIV:0711.2534;%%

%\cite{Haba:2007rj}
\bibitem{HM}
  N.~Haba and N.~Maru,
  ``A Simple Model of Direct Gauge Mediation of Metastable Supersymmetry
  Breaking,''
  arXiv:0709.2945 [hep-ph].
  %%CITATION = ARXIV:0709.2945;%%

%\cite{Xu:2007az}
\bibitem{XY}
  F.~Xu and J.~M.~Yang,
  ``An Extension for Direct Gauge Mediation of Metastable Supersymmetry
  %Breaking,''
  arXiv:0712.4111 [hep-ph].
  %%CITATION = ARXIV:0712.4111;%%

\end{thebibliography}
\end{document}